# Probing the Universe's baryons with fast radio bursts

*Fast radio bursts were discovered just over a decade ago, and their origin remains a mystery. Despite this, astronomers have been using them to investigate the matter through which their bright, impulsive radiation travels.*

Jean-Pierre Macquart

As the first impulsive radio transients to be detected beyond the realms of our local Galaxy Group, Fast Radio Bursts (FRBs) offer diagnostics that could revolutionise studies of the ordinary, baryonic, matter in the Universe. If located, as supposed, at cosmological distances, they hold the capacity to detect and examine in detail the elusive 30–80% of baryonic matter resident in the tenuous ~$10^{5-7}$ K warm–hot component of the Intergalactic Medium (IGM), examine the magnitude of ill-constrained baryonic feedback processes in galaxies, and possibly even yield new insights into the nature of dark energy. The burst properties may also contain signatures of their circumburst environment, divulging the identity of these mysterious events and the cause of their ultra-luminous radio emission.

A defining characteristic of an FRB pulse is that it is so strongly dispersed that its radiation must have propagated through an amount of ionised matter well beyond that attributable to the interstellar medium (ISM) of our Galaxy. Dispersion retards the pulse arrival time by an amount that is quadratic with wavelength, and proportional to the column density of ionized matter present between the burst and the Earth. Contributions to the dispersion measure (DM, expressed in pc cm$^{-3}$) of an FRB come from four regions: the plasma that pervades the Milky Way (MW), intergalactic space (the IGM), that of an FRB host galaxy, and any non-relativistic matter in the immediate burst environment.

The MW contribution is considerably smaller than the DMs of FRBs reported so far, which span the range 114–2,596 pc cm$^{-3}$. The interstellar medium confined to the MW disk contributes typically 30 pc cm$^{-3}$ at Galactic latitudes |*b*| > 30˚. However, FRBs also propagate through an additional highly uncertain 15–50 pc cm$^{-3}$ from the Galactic halo. The lowest DMs exhibited by FRBs provide a means to deduce the total baryonic mass of the Galactic halo and its profile.

From the comparatively small contributions to the DM of an FRB by Galactic material, it is clear that most FRB DMs will be dominated by the extragalactic contribution.

**Intergalactic baryons & Galactic Feedback**
The outstanding lure of FRBs lies in their potential to weigh the baryonic content of the Universe. They thus directly address the amount and location of all the ordinary nonexotic (baryonic) matter believed "missing" in the present Universe[1]. Until very recently, censuses of the nearby Universe fail to account for roughly half of the entire baryonic matter content that is estimated to exist on the basis of both cosmological theory and measurements of the hydrogen density in intergalactic gas 10 billion years ago[1,2]. FRBs are an exquisitely high-precision probe of the missing baryons because the effect of dispersion accounts for every single ionized baryon that lies between the burst and the Earth. This enables their detection in a fundamentally new way, in contrast with other techniques, chiefly absorption spectroscopy, which is limited by telescope sensitivity and uncertainty in the abundance of the absorbing element.

The IGM makes a large, and possibly dominant, contribution to the DM if these bursts are located at cosmological distances. Over cosmological distances both the continuous redshifting of the radiation and the curvature of the Universe disrupt the one-to-one correspondence between the plasma column and the DM that is deduced from the quadratic proportionality between the pulse arrival time and wavelength. The relationship between the expected mean intergalactic contribution to the DM and the burst redshift, z, encapsulates much of the utility of FRBs as probes of intergalactic space:

$$\langle \mathrm{DM_{IGM}} \rangle = \frac{3cH_0\Omega_\mathrm{b}}{8\pi Gm_p} \int_0^z \frac{(1+z')f_\mathrm{IGM}\left(\frac{3}{4}X_\mathrm{e,H}(z') + \frac{1}{8}X_\mathrm{e,He}(z')\right)}{\sqrt{\Omega_\mathrm{m}(1+z')^3 + \Omega_\Lambda f(z',w)}} dz' \quad (1)$$

where a flat Universe is assumed, where $\Omega_\mathrm{b}$, $\Omega_\mathrm{m}$ and $\Omega_\Lambda$ are the baryonic, matter and dark energy densities relative to the critical density, $H_0$ is the Hubble constant, $f_\mathrm{IGM}$ is the fraction of all baryons residing in the IGM, and *f(z',w)* ≈ 1 if the equation of state parameter of dark energy *w* ≈ *-1*. The ionization fractions of hydrogen and helium are denoted by $X_\mathrm{e,H}$ and $X_\mathrm{e,He}$ respectively.

The baryonic matter density can be deduced directly from the constant of proportionality between the measured DM and the burst host galaxy redshifts. The potential of FRBs as cosmic weighbridges will be realised only in the era of routine FRB host galaxy identifications. Considerable stochasticity is expected in the IGM signal, so that the measurement of quantities in equation (1) will require detections of an ensemble of bursts. Characterisation of the host galaxies of FRBs would also help constrain the likely DM contribution due to the host galaxy ISM. Cosmological simulations[2] place reasonable bounds on the expected signal variance in the IGM, and thus the number of bursts required to attain a given level of certainty.

Far from a hindrance, stochasticity in the DM-redshift relation, equation (1), provides an entirely new diagnostic of baryonic feedback from galaxies. Inhomogeneity in the IGM DM signal is primarily caused by the random intersection of galaxy halos with the line of sight. The DM distribution is a sensitive probe of the nature of the galactic baryonic halos, whose extents and masses are governed by the feedback processes that expel the baryons into these structures. Cosmological simulations[2] have shown that the masses of galaxies that possess appreciable halos and their sizes directly relate to the shape and variance of the DM distribution. The greater the strength of feedback — the larger the halos — the more halos will intersect the line of sight.

**The global tapestry of the Universe**
Helium contributes up to 14% of the total electron density of the IGM, and FRBs detectable at some redshift beyond *z*~3 will bear the imprint associated with the epoch of helium reionization. The detection and characterisation of this epoch may hold important lessons in searches for the signature of the epoch of hydrogen reionization, at yet higher redshifts, by illustrating how quickly and homogeneously such a phase transition occurs in the Universe. The signature of helium reionization should be manifest in the distribution of observed FRB DMs[3]. Thus its detection does not necessitate distance measurements of large numbers of FRBs, but merely their detection, which is much easier. Nonetheless, the ultimate detectability of this signal depends how on how rapidly helium reionization occurs, the number of FRBs detectable at such redshifts, and the amplitude of the scatter in the DM-redshift relation.

A yet vastly more ambitious target is to use the DM-redshift relation to measure the curvature of the Universe and hence place limits on the nature of dark matter through a measurement of $w$, the equation of state of dark energy, or even its redshift dependence. This would require measurement of the distances to a sample of at least many hundreds, perhaps thousands, of FRBs to average down stochasticity in the DM, and it is presently unclear to what extent systematic effects in the DM-redshift relation could further swamp this signal.

The epitome of FRB dispersion measure science would be a detailed tomographic reconstruction of the entire cosmic web. A reasonable minimum requirement for a such an undertaking would involve the detection at least one FRB on average per void. This may be within the reach of upcoming wide-field survey telescopes. To put this in perspective, a 30 deg$^2$ patch of sky out to $z$=0.5 contains ~2,000 voids. A survey using a telescope with the sensitivity of the Parkes radio telescope nets events at a rate of 3,000 events over the sky per day (down to a fluence of 2 Jy ms), or 800 events per year over 30 deg$^2$.

**The host galaxy and burst environment**
The DMs of FRBs could also contain significant contributions from their host galaxies or plasma associated with the burst environment. Simple assumptions place reasonable bounds on the host galaxy contribution. For instance, an event that occurs in the disk of a spiral galaxy similar to the Milky Way would have a typical integrated ISM contribution of 30 pc cm$^{-3}$ for all |$b$| > 10° (that is, over 83% of its solid angle), with a possible further 15–30 pc cm$^{-3}$ from its halo. Higher contributions are possible for special viewing geometries, for more massive gas-rich galaxies (e.g. some classes of elliptical galaxies), or if the bursts predominately occur in special regions. For instance, some have posited ISM contributions up to ~10$^3$ pc cm$^{-3}$ if the bursts originate in the dense gaseous regions within the inner pc of their host galaxies[4].

The DM component directly associated with the circumburst environment is a powerful discriminant between competing burst progenitor hypotheses. Large DM contributions are expected in some progenitor models (for example, young neutron stars embedded in a supernova remnant, models involving copious pre-burst stellar mass loss), while in others the system is expected to be evacuated prior to burst (e.g. binary black hole and neutron star mergers). If the circumburst medium dominates the total dispersion measures of FRBs, rather than intergalactic baryons, it would cast doubt on the supposition that the bursts are at large cosmological distances. The debate on the circumburst environment contribution is coupled directly to arguments about the luminosities and energetics of the FRB phenomenon; DMs comprising at most a 10% intergalactic contribution would imply a population that is extragalactic but still largely confined to the nearby, $z$~0.1 (500 Mpc) Universe. This would reduce the inferred energetics of FRBs by at least two orders of magnitude relative to a population bounded at a redshift $z$~1.

Although the contribution of the circumburst medium is hard to quantify a priori, it immediately implies a number of testable predictions. A dense medium cocooning a burst would manifest its presence by absorbing low frequency (<300 MHz) radio emission due to free-free absorption. This is a possible explanation of the dearth of FRB detections by new-generation low frequency arrays, notably the Murchison Widefield Array (MWA) and the LOw Frequency ARray (LOFAR).

A potent means of assessing the contribution of the circumburst and host galaxy medium relative to the IGM is based on the statistical relation between DM and brightness: evidence of a clear DM-brightness relation, albeit with a large expected scatter[3], would be a strong indication that a large component of the DM is related to its distance, and hence originates from intergalactic baryons.

Two other notable effects provide ancillary diagnostics of the plasma along the line of sight: Faraday rotation, and temporal smearing due to plasma inhomogeneity (turbulence).  The effect of Faraday rotation, a rotation of the plane of linear polarization resulting from propagation through a magnetised plasma, measures the integrated product of the electron density and the magnetic field along the direction of propagation.  Extremely high (>80%) levels of linear polarization are detected in some FRBs, rendering this effect readily measurable.

Faraday rotation proffers a means to open up studies of the magnetic fields embedded in intergalactic space. In principle, the intergalactic magnetic field could be gleaned from a sample of FRBs sufficiently large to establish the correlation between Faraday rotation and DM, once enough is known about FRB host galaxies to reveal the fraction that is intergalactic in origin. This technique has the virtue over Faraday rotation studies of other cosmological populations in that it obviates the need to obtain their distances (redshifts) independently.

The magnetic field strengths expected may make the intergalactic increment small compared to other contributions. Indeed, the primary immediate utility of Faraday rotation likely lies in its diagnosis of the circumburst medium.  The rotation measure of a dense, thick circumburst medium could dominate the contributions from the ISM of our Galaxy and the host galaxy, the latter both typically tens of radians per square metre.  This is especially so since one would generically expect, based on equipartition arguments, higher magnetic field strengths in regions of high density.

Some ~10% of all hitherto detected FRBs show a pulse profile whose width increases as $\lambda^4$, a signature of temporal smearing due to multipath propagation, and a tell-tale sign that the radiation encountered a turbulent region along the line of sight.  The temporal smearing does not originate in our Galaxy; the measured smearing times are on the order of milliseconds but the Galactic contribution is at most a few microseconds for bursts observed at high Galactic latitudes.  It is unclear whether the scattering originates in a circumburst medium, the ISM of the host galaxy, or in the tenuous halos of intervening systems embedded in the IGM.  The latter is an intriguing possibility, given that there is strong lever-arm that up-weights the contribution of matter midway along the line of sight relative to matter close to the burst: the effect is roughly proportional to the distance between the burst and plasma.  Whether this is relevant depends on the likelihood of intersecting the halo of an intervening system, and density of the matter therein: estimates[2] suggest this effect is likely to be important.  If so, the effect of scattering could be intertwined with the subject of feedback and DM contributions from discrete numbers of intervening systems. Some observations (see, for example, ref. 5) have placed limits on the distance to the scattering for some bursts, but they are relatively unconstraining so far.

**Outlook**
There is now strong evidence that FRB dispersion measures contain a significant contribution from intergalactic baryons.  FRBs do show a dispersion measure-brightness relation[6].  The populations detected by Parkes[7] and ASKAP[6] are sufficiently large to examine this relation, showing that the average DM of Parkes events is twice as high as the corresponding average over ASKAP events.  The two surveys differ in limiting fluence by a factor of ~20.  This result shows that the DM encodes information on the distance of these bursts, and thus that the IGM contributes substantially. The cosmological nature of the population is corroborated by the burst brightness distribution (the source counts), which differs from a Euclidean slope at the >90% confidence level[6].

The Faraday rotation measures of FRBs lend credence to the conjecture that the phenomenon encompasses two physically distinct events types. Where linear polarization has been detected, several FRBs have rotation measures largely consistent only with the contribution of our Galaxy's ISM, while a few[8,9] have somewhat higher magnitude RMs (~200 rad m$^{-2}$). However, the RM of the repeating FRB is exceptional, with a rotation measure of ~$10^5$ rad m$^{-2}$ that also appears to be variable[10]. That the only known repeating FRB also possesses an extraordinarily high rotation measure, and has typical burst fluences orders of magnitude lower than those single-burst events found by Parkes and ASKAP, suggests that FRB taxonomy may soon be due a revision.

The ultimate utility of FRBs as cosmic probes rests on the nature of the bursts themselves, through their luminosity distribution and the separability of propagation effects in the burst environment from those contributed by intergalactic baryons. If the FRB luminosity function is sufficiently shallow, their reduction in brightness with distance is outweighed by the increase in the number of bright bursts within the search volume, and the bursts should be easily detectable over a large span of cosmic time. Of course, this is true for many other astrophysical populations. The brightest radio active galactic nuclei are also among the most distant. Gamma Ray Bursts are detected beyond a redshift of $z$=6. What is different here is that FRBs are susceptible to a whole suite of supremely sensitive propagation effects like no other cosmological population, and even come with an inbuilt independent measure of distance.

*Jean-Pierre Macquart is at the International Centre for Radio Astronomy Research (ICRAR), Curtin University, Perth, Australia.*
*Email: J.Macquart@curtin.edu.au*